\documentclass[twocolumn,showpacs,preprintnumbers,amsmath,amssymb,floatfix]{revtex4}
\usepackage{graphicx}
\usepackage{epsfig}
\usepackage{amsfonts}
\usepackage{amssymb}
\usepackage{epsf}
\usepackage{color}
\newcommand{\insertplot}[5]{\begin{figure}
 \hfill\hbox to 0.05in{\vbox to #5in{\vfill
 \inputplot{#1}{#4}{#5}}\hfill}
 \hfill\vspace{-.1in}
 \caption{#2}\label{#3}
 \end{figure}}
\newcommand{\inputplot}[3]{
 \special{ps: plotfile #1}
}

\newcounter{fig}

\def\beq{\begin{equation}}
\def\eeq{\end{equation}}
\def\bea{\begin{eqnarray}}
\def\eea{\end{eqnarray}}

\begin{document}

\title{Gauss-Bonnet Inflation}

\author{ 
{\bf Panagiota Kanti}
}
\affiliation{ 
Department of Physics,
University of Ioannina, Ioannina GR-45110, Greece,}
\smallskip
\affiliation{
Department of Physics, National and Kapodistrian University of Athens,
Athens, Greece}

\author{ 
{\bf Radouane Gannouji}
}
\affiliation{Instituto de F\'{\i}sica, Pontificia Universidad  Cat\'olica de Valpara\'{\i}so, 
Casilla 4950, Valpara\'{\i}so, Chile}

\author{ 
{\bf Naresh Dadhich}
}
\affiliation{Centre for Theoretical Physics, Jamia Millia Islamia, New Delhi-110025, India}
\affiliation{Inter-University Centre for Astronomy \& Astrophysics, Post Bag 4, Pune 411 007, India}

{}
\date{\today}
\begin{abstract}
We consider an Einstein-Scalar-Gauss-Bonnet gravitational theory, and argue
that at early times the Ricci scalar can be safely ignored. We then demonstrate that
the pure scalar-Gauss-Bonnet theory, with a quadratic coupling function, naturally
supports inflationary -- de Sitter -- solutions. During inflation, the scalar
field decays exponentially and its effective potential remains always bounded.
The theory contains also solutions where these de Sitter phases possess a
natural exit mechanism and are replaced by linearly expanding -- Milne -- phases.
\end{abstract}
\maketitle

\section{Introduction}

Einstein's theory of gravity, based on General Relativity, is known to possess
many successes but also many open questions. In the context of Cosmology,
the ``old problems'' of flatness, horizon, monopoles and density perturbations
were solved by the introduction of inflation, a very fast expanding, early phase
in the history of our universe \cite{Guth, Linde, Starobinsky}. New questions
have however emerged, such as the nature of dark matter and dark energy
or the coincidence problem, which are still persisting despite all the efforts.

Searching for alternative and maybe more fundamental theories of Gravity,
that would allow for new directions of thinking and for novel solutions,
has been a path followed by many scientists. Looking at higher dimensions
for inspiration, the Lovelock theory \cite{Lovelock} emerges as a natural
higher-dimensional ($d>4$) generalization: its 
action is a homogeneous polynomial of degree $N$ of Rie\-mann curvature,
with Einstein's term $R$ arising for $N=1$, the quadratic Gauss-Bonnet term 
$R^2_{\rm GB} = R_{\mu\nu\rho\sigma} R^{\mu\nu\rho\sigma}
- 4 R_{\mu\nu} R^{\mu\nu} + R^2$ for $N=2$, and so on. The most remarkable 
and distinguishing basic feature of the theory is that the ensuing field equations are of 
second order as in Einstein's gravity, a requirement necessary to ward off
undesirable features like ghosts. 

Our primary candidate for a generalised gravitational theory in four dimensions
is the one given by the heterotic superstring effective action \cite{Zwiebach:1985uq,Gross, Tseytlin},
where the Gauss-Bonnet (GB) term appears as an $\alpha'$-order correction. 
In addition, this term is the highest, non-zero one in the Lovelock polynomial in
$d=4$ -- in $d$ dimensions, only the terms with $N \leq d/2$ are non-trivial
\cite{Lovelock}. But being a topological invariant in 4-dimensions, the GB term, in
the context of any theory, should always be coupled to a scalar field to get a
non-trivial contribution to the field equations. 
The presence of the GB term in the string effective theory,
or in more generalised string-inspired ones, has been shown to lead to novel
solutions such as singularity-free cosmological solutions \cite{Antoniadis, KRT},
hairy black holes \cite{KMRTW, Torii} or even traversable wormholes \cite{KKK}.

In all cases, the GB term was shown to be the dominant agent that allowed
for the existence of these solutions. Motivated by this, here we consider an
Einstein-scalar-GB theory with a polynomial coupling function between the 
scalar field and the GB term. We focus our analysis on the early-time
evolution of the universe and argue that the Ricci scalar is again sub-dominant
to the GB term, and thus it can be ignored. We then demonstrate that analytical,
elegant, inflationary solutions with attractive characteristics -- compared to
the existing models -- are naturally supported by the field equations.

\section{The Einstein-Scalar-GB Theory}

We consider the following gravitational theory of a scalar field $\phi$
\begin{eqnarray}  
{\cal S}=\int d^4x \sqrt{-g} \left[\frac{R}{2\kappa^2}-\frac{(\nabla \phi)^2}{2} +
\frac{1}{8}\,f(\phi) R^2_{\rm GB}\right],
\label{act}
\end{eqnarray} 
coupled non-minimally to gravity via the Gauss-Bonnet term $R^2_{\rm GB}$
by a general coupling function $f(\phi)$. 
The variation of the action (\ref{act}) with respect to the metric tensor
and scalar field leads to the field equations
\beq
R_{\mu\nu}-\frac{1}{2}g_{\mu\nu}R + P_{\mu \alpha \nu \beta}\nabla^{\alpha \beta}f
=\partial_\mu\phi \partial_\nu \phi
- g_{\mu \nu}\,\frac{(\nabla\phi)^2}{2}\,,
\eeq
\beq
\frac{1}{\sqrt{-g}}\,\partial_\mu\left[\sqrt{-g}\,\partial^\mu \phi \right] +
\frac{1}{8}\,f' R^2_{GB}=0\,,
\label{dilaton-eq-cov}
\eeq
where $f' \equiv df/d\phi$, the constant $\kappa^2 \equiv 8\pi G$ has been set
to unity, and $P_{\mu\alpha\nu\beta}$ is defined as 
\beq
P_{\mu \alpha \nu \beta}=R_{\mu \alpha \nu \beta}+2g_{\mu[\beta}R_{\nu]\alpha}+
2g_{\alpha[\nu}R_{\beta]\mu}+Rg_{\mu[\nu}g_{\beta]\alpha}\,.
\eeq

We assume that the line-element has the Friedmann-Lema\^itre-Robertson-Walker form
\beq
ds^2 =-dt^2 +a^2(t)\left[\frac{dr^2}{1-kr^2}+r^2 (d\theta^2 + 
\sin^2\theta d\varphi^2)\right],
\label{metric}
\eeq
with $k=0, \pm 1$ denoting the curvature of the 3-dimensional space.
Then, the field equations take the explicit form
\begin{eqnarray}
\hspace*{-0.5cm} 3(1+\dot f H)\bigl(H^2 +\frac{k}{a^2}\bigr) &=& \frac{\dot \phi^2}{2}\,,
\label{Ein-1}\\[1mm]
2(1+H\dot f) (H^2+\dot H) \hspace*{1.5cm}
&&\nonumber \\ 
+(1+\ddot f) \bigl(H^2+
\frac{k}{a^2}\bigr)&=&-\frac{\dot \phi^2}{2}\,, \label{Ein-2}\\[1mm]
\ddot \phi+3 H \dot \phi-3f' \bigl(H^2+\frac{k}{a^2}\bigr)
(H^2+\dot H)&=&0\,, \label{dilaton-eq-0}
\end{eqnarray}
where $H \equiv \dot a/a$ is the Hubble parameter and the dot denotes
a derivative with respect to time. In this work, we will consider a polynomial
coupling function, i.e. $f(\phi)=\lambda\,\phi^n$, where $\lambda$ is a
constant and $n$ an integer. The case with $n=0$ is equivalent to ignoring
the GB term; the cases with $n=1$ and $n=2$ will be studied in Sections
III and IV, respectively.


\section{The Role of the Ricci Scalar }

In this section, we will formulate an analytical argument to demonstrate
that, in the strong-gravity regime, the presence of the Ricci scalar significantly 
complicates the derivation of a cosmological solution without adding any
important effect to its dynamics.  Here, we will consider the case of $n=1$, for
which the scalar equation (\ref{dilaton-eq-0}) can be 
integrated once to yield the relation
\beq
\dot \phi=\frac{D}{a^3} + \frac{\lambda \dot a\,(3k+\dot a^2)}{a^3}\,,
\label{rel-phi-R}
\eeq
with $D$ an integration constant. The second term in the above relation is
clearly the one associated to the GB term, and we will focus on this, setting $D=0$. 

Combining Eqs. (\ref{Ein-1})-(\ref{dilaton-eq-0}) and using that $f=\lambda \phi$,
we obtain the constraint
\bea
&&  \hspace*{-1cm}
4\bigl(H^2+ \frac{k}{a^2}\bigr) +2(H^2+ \dot H)\,(1+\lambda H \dot \phi)\nonumber\\
&&  \hspace*{1cm}+ 
3\lambda^2(H^2 +\dot H) \bigl(H^2+ \frac{k}{a^2}\bigr)^2=0\,.
\label{Eq1and2-R-1}
\eea
If we substitute $\dot\phi$ via Eq. (\ref{rel-phi-R}), the above equation has no
dependence on the scalar field. After some algebra, it can be integrated once to give
\beq
3a^4=-\lambda^2 \left(\frac{13 k \dot a^2}{2} +\frac{5 \dot a^4}{2}
+\frac{2 k^3}{k+\dot a^2}\right) + c_1\,,
\label{adot-linear}
\eeq
with $c_1$ an integration constant again. For $k \neq 0$, the above 
cannot be easily solved for $\dot a$. However, for a flat universe ($k=0$),
we easily find that
\beq
\dot a ^4=\frac{2c_1}{5\lambda^2}\,\left(1-\frac{3 a^4}{c_1}\right).
\eeq
The solution of the above equation can be written in the non-explicit form 
\beq
a(t)\,F\left[\frac{1}{4},\frac{1}{4},\frac{5}{4};\frac{3 a^4(t)}{c_1}\right]=
\left(\frac{2c_1}{5\lambda^2}\right)^{1/4}(t+t_0)\,,
\eeq
where $F(a,b,c;x)$ is the hypergeometric function. The above describes an
increasing scale factor with time 
that is however bounded
from above due to the convergence criterion $a(t) \leq a_{max}=(c_1/3)^{1/4}$
of $F$. In the limit $a \rightarrow 0$, i.e. in the 
strong-gravity limit, the relation between the scale factor and the time coordinate
becomes linear with a Big-Bang singularity emerging at a finite value of time. 

Let us now focus on the strong-gravity regime from the beginning and
assume that, in that limit, the Ricci scalar $R$ will be subdominant to
the GB term. In this case, the scalar equation (\ref{dilaton-eq-0})
remains the same, while the two gravitational equations will be modified
by loosing all unity terms appearing inside brackets on their left-hand-sides.
Then, Eq. (\ref{Ein-1}) can be solved for $\dot \phi$, even for a general
coupling function $f(\phi)$, to give
\beq
\dot \phi=6f'H \bigl(H^2+\frac{k}{a^2}\bigr)\,.
\label{phi-dot}
\eeq
Following a similar analysis as above, i.e. combining
Eqs. (\ref{Ein-1})-(\ref{dilaton-eq-0}), employing Eq. (\ref{phi-dot})
and the fact that $f=\lambda\phi$, we now arrive at the simpler constraint
\beq
\Bigl(5H^2+\frac{k}{a^2}\Bigr)\,(H^2+\dot H) =0\,.
\label{Eq1and2-v1}
\eeq
The general solution of the above differential equation is: $a(t)=A t+B$,
describing a linearly-expanding universe at early times. This is in full
agreement with the result obtained above in the context of the full analysis
and in the early-time limit. By plotting the two solutions \cite{DGK2},
one may easily see that these perfectly match at early times,
while a deviation begins to appear, for a given $\lambda$, the sooner
the larger the value of the integration constant $c_1$ is. 
A similar conclusion may be reached for the behaviour of the scalar field:
using either Eq. (\ref{rel-phi-R}) or (\ref{phi-dot}) and taking the limit
of early times,  we obtain $\phi \sim (At+B)^{-2}$ with a divergence
appearing at the initial singularity.


\section{Inflationary Expansion}

In this section, we will proceed to study the case of a quadratic coupling
function, $f=\lambda \phi^2$. Unfortunately, in this case, the set of
equations is significantly more complicated and the formulation of a similar
analytic argument regarding the role of the Ricci scalar is impossible. However,
since the coupling function determines the weight of the GB term in the theory,
we expect that its exact form merely determines the point in time where the
GB term begins to dominate over the Ricci scalar in the early-time,
strong-gravity regime. Thus, here we will ignore again the Ricci scalar and
focus on an appropriately chosen early-time period.
Equation (\ref{phi-dot}) still holds, however, the different form of the coupling
function modifies the constraint (\ref{Eq1and2-v1}) as follows
\beq
\bigl(5 H^2+\frac{k}{a^2}\bigr)(H^2+\dot H) +
24 \lambda H^2 \Bigl(H^2+\frac{k}{a^2}\Bigr)^2=0\,.
\label{constraint}
\eeq
The above differential equation involves only the scale factor $a(t)$, 
however it is only for $k=0$ that it can be integrated twice to yield
an explicit solution -- for $k \neq 0$, a transcendental equation arises
after the first integration \cite{DGK2}. Thus, for a flat space, we have
\beq
\dot H+H^2(1-\frac{H^2}{H_{dS}^2})=0\,,
\label{cons-k0}
\eeq
where we have defined $H_{dS}^2=-5/24\lambda$. The above implies that a
de Sitter solution, with $H$ a constant and thus
$\dot H=0$, exists when $\lambda<0$ describing an inflationary phase. 
The corresponding accelerating solution with $H_{dS}^2=-5/24\lambda$ is a repeller,
while the second solution $H=0$, describing a Minkowski spacetime, is an attractor.
Therefore, in order to have a natural exit from the inflationary phase
we have to consider an initial condition ($H_i$) such that $H_i<H_{dS}$.
Note that if we consider $H_i > H_{dS}$, the Universe will be eternally inflating $(\dot H>0)$.
Integrating Eq. (\ref{cons-k0}) once, we obtain
\beq
H=\frac{H_{dS}}{\sqrt{1-\frac{C_1}{12\lambda}\,a^2}}\,,
\label{general}
\eeq
where $C_1$ is an arbitrary integration constant. Depending on the values of
the two parameters $\lambda$ and $C_1$, the above equation results to
different types of cosmological solutions. The existence of de Sitter implies
$\lambda<0$ and the natural exit implies $C_1\ge 0$ ($H<H_{dS}$). Here, we
will present only these two cases that support a non-eternal inflationary phase
of expansion for the universe - for a more comprehensive analysis,
please refer to \cite{DGK2}.


\subsection{The case with $C_1=0$ and $\lambda<0$}

If $C_1=0$, then integrating once more Eq. (\ref{general}), we find 
\beq
a(t)=a_0\,\exp\left(\sqrt{\frac{5}{24 |\lambda|}}\,t\right)\,.
\label{sol-C10}
\eeq
The profile of the scale factor is shown in Fig. 1. We observe that the smaller
the value of the coupling constant $|\lambda|$ is, the faster $a(t)$ evolves with time.
The above solution may be considered as an alternative to the usual 
inflationary solutions following from the Einstein-Hilbert action which require 
an appropriate potential for the scalar field -- here, it is the GB term itself
that provides the necessary potential.

\begin{figure}[t]
  \begin{center}
\includegraphics[width = 0.42 \textwidth] {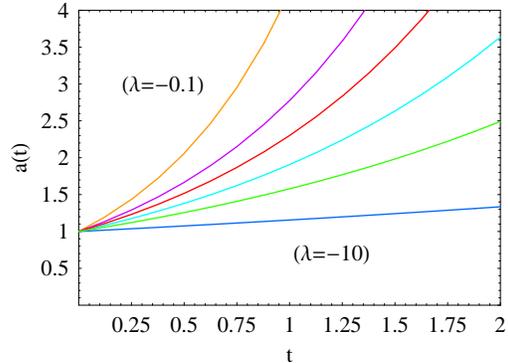}
    \caption{The scale factor as a function of time for $C_1=0$ and
	$|\lambda|=0.1, 0.2, 0.5, 1, 2, 10$ (from top to bottom).}
   \label{inflation-scale}
  \end{center}
\end{figure}

The solution for the scalar field can be found via Eq. (\ref{phi-dot}): 
for $k=0$ and $f'=2\lambda\phi$, it can readily be integrated to 
give the solution
\beq
\phi(t)=\phi_0\,\exp\left(-\frac{5}{4}\,\sqrt{\frac{5}{6|\lambda|}}\,t\right)\,.
\label{solphi-C10}
\eeq
The above describes a regular, exponentially decaying scalar field.
One may easily verify that the combination of solutions (\ref{sol-C10}) and
(\ref{solphi-C10}) satisfies the complete set of field equations.


The scalar field starts from an arbitrary value $\phi_0$ and quickly reduces
to zero -- again, the smaller the value of $|\lambda|$ is, the faster it decays.
As a result, the coupling function $f(\phi)=\lambda \phi^2$ remains always
bounded. The effective potential of the scalar field receives contributions from
both the GB term and the coupling function and has the explicit form
\beq
V_{eff} \equiv -\frac{1}{8}\,f(\phi) R^2_{\rm GB}=
\frac{25}{24}\,\frac{\phi^2}{8 |\lambda|}\,. \label{Veff}
\eeq
Therefore, $V_{eff}$ also remains bounded as the field evolves; at the same time,
it may become arbitrarily large at early times by appropriately choosing the value of
the coupling constant $\lambda$. This is in sharp contrast to the behavior seen
in more traditional inflationary models, such as the chaotic \cite{Linde}, where
the effective potential blows up for the super-Planckian initial values of the field 
unless a fine-tuning is imposed on the parameters of the model -- as we will
 see in Sec. IV, such an unnatural initial value is not needed in our case for the
necessary number of e-foldings to be obtained. It is also
a significant improvement compared to models for inflation
\cite{Starobinsky} where the scalar potential remains bounded but its value
is too small to justify its dominance over any other distribution of
matter in the universe and thus allow inflation to be realised.


\begin{figure}[t]
  \begin{center}
\includegraphics[width = 0.37 \textwidth] {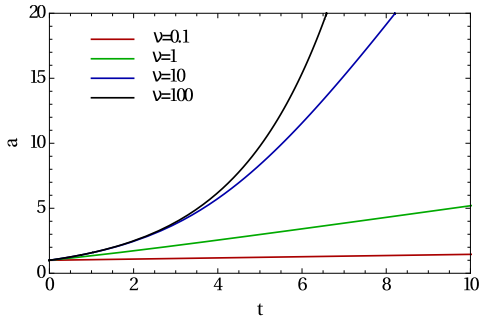}
    \caption{Cosmological solutions with $C_1>0$, $\lambda<0$, and for
	$\nu=0.1,1,10,100$.}
   \label{complete}
  \end{center}
\end{figure}

\subsection{The case with $C_1>0$ and $\lambda<0$}

We will now demonstrate that the pure de Sitter solution
of the previous case is the early-time limit of a more interesting class
of cosmological solutions that follow when $\lambda<0$ and $C_1$ positive.
In this case, by using the change of variable $a=\nu\,\tan w$, where 
$\nu^2 \equiv 12 |\lambda|/C_1$, Eq. (\ref{general}) can be again
integrated to yield
\beq
\sqrt{a^2+\nu^2} +\nu \ln\left(\frac{\sqrt{a^2+\nu^2}-
 \nu}{a} \right) = \sqrt{\frac{5}{2 C_1}}\,(t+t_0)\,.
\label{sol-lam-neg}
\eeq
Plotting the above (see Fig. 2), the scale factor is found to be an increasing function
of time. In the limit $a \rightarrow 0$, i.e. at early times,
Eq. (\ref{sol-lam-neg}) reduces to the pure de Sitter solution (\ref{sol-C10});
on the other hand, expanding (\ref{sol-lam-neg}) for  
$a^2 \gg \nu^2$, we obtain a linear dependence of the scale factor
on time, $a(t) \simeq \sqrt{5/2C_1}\,t$. Thus, in this case
the universe naturally interpolates between a de Sitter phase at early times
and a Milne phase at later times.  

In order to find the solution for $\phi$, we observe that, for $f(\phi)=\lambda \phi^2$,
it holds that $f''=f'/\phi$. Then, using Eq. (\ref{phi-dot}) in Eq. (\ref{constraint})
and integrating once, we obtain the relation $\phi^2 \dot a^5=const. \equiv C_0$.
Then, the following implicit expression may be derived for the scalar field
\beq
\phi^2=C_0\,\left(\frac{2C_1}{5}\right)^{5/2}\frac{(a^2+\nu^2)^{5/2}}{a^5}\,.
\eeq
For early times, $\phi$ is again decreasing exponentially while, at later times,
it reduces to a constant. Its effective potential is also decreasing, exponentially at
early times and as ${\cal O}(t^{-6})$ at later times, remaining again bounded.
However, we expect that the late-time behaviour of both the scale factor and the
scalar field will start to deviate from the above expressions due to the increasing role of the
Ricci scalar as the universe expands and to the contribution of radiation/matter.
Only a numerical analysis of the full model could reveal the duration of the
Milne phase and its implications for late cosmology as well as the fate of
the scalar field.

The early de Sitter phase should have a long-enough duration in order to
resolve the problems of the standard cosmology. Using the form (\ref{general}),
we demand the initial condition to satisfy $H_i\simeq H_{dS} (1-e^{-2N})$ which
implies $C_1\simeq |\lambda| e^{-2N}$, where $N$ is the number of e-foldings $(N>60)$.
According to this, the smaller the value of $C_1$, the longer the acceleration; this was
expected since for $C_1=0$, the acceleration is eternal.
Therefore the number of e-foldings during inflation depends on the
initial condition $H_i$.

\begin{figure}[t]
  \begin{center}
\includegraphics[width = 0.37 \textwidth] {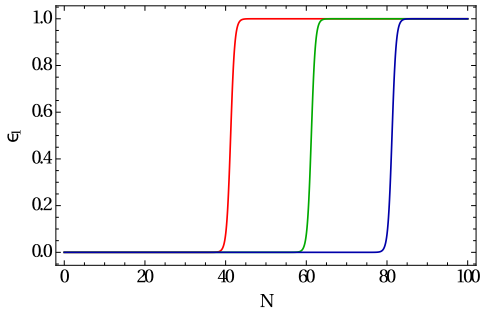}
    \caption{Evolution of the slow-roll parameter $\epsilon_1$ as function of
	$N=log(a)$ for different initial conditions $C_1= |\lambda| e^{-2N}$ 
which gives an e-folding of $(40,60,80)$, respectively.}
   \label{slow-roll}
  \end{center}
\end{figure}

We may finally define the slow-roll parameters $\epsilon_1 \equiv -\dot H/H^2$ and
$\epsilon_{i+1} \equiv -\dot \epsilon_i/H \epsilon_i$ as follows
\begin{align}
\epsilon_{2n+1} &= \epsilon_{1} =1-H^2/H_{dS}^2\,,\\
\epsilon_{2n} &= \epsilon_{2} =2 H^2/H_{dS}^2\,.
\end{align}
During inflation, the odd slow-roll parameters approach zero, $\epsilon_{2n-1} \simeq 0$,
while the even parameters are of order unity, $\epsilon_{2n}\simeq 2$. The slow-roll
approximation is thus violated. A very similar behavior exists in the so called ultra-slow
roll inflation \cite{Kinney:2005vj} (also known as fast-roll inflation \cite{Motohashi:2014ppa}). 

At later times, the slow-roll parameters move away from the above values to mark
the end of inflation. In Fig. \ref{slow-roll}, we depict the evolution
of $\epsilon_1$, for different initial
conditions: inflation is realized at early times, when $\epsilon_1 \simeq 0$,
and a natural exit takes place at later times as $\epsilon_1$ moves to values
closer to unity.

Let us finally note that the cases of a general polynomial coupling function,
i.e. with $f(\phi)=\lambda \phi^n$ and $n$ an arbitrary integer, as well as that
of an exponential coupling function, i.e. with $f(\phi)=\lambda e^{\kappa \phi}$ and
$\kappa$ an arbitrary number, were also studied \cite{DGK2}. In the first case, it was
demonstrated that de Sitter-type solutions, either as exact or limiting solutions, arise
only for the case of $n=2$. In the latter case, it was shown that a scalar-GB theory
with an exponential coupling function admits, at early times, only slowly expanding
solutions of the form $a(t) \simeq (A t +B)^{1/5}$. Therefore, the particular model
apparently singles out the quadratic coupling function as the unique choice that
allows for the emergence of inflationary solutions or solutions with additional
attractive characteristics from the cosmological point of view \cite{DGK2}.


\section{Conclusions}

We have considered an Einstein-Scalar-GB gravitational theory,
and developed an analytical argument to show that the Ricci scalar can be
safely ignored when we focus on early times. Note that in the high energy regime of
the early Universe, higher-order curvature terms would naturally be the most pertinent.
Here,  we have considered only those terms inspired by the Lovelock expansion, 
that lead to second-order differential equations, a necessary condition to ward
off ghost fields from the theory. As the higher Lovelock polynomials, with $N \geq 3$,
are trivial in four dimensions, even if coupled to a scalar field (since the Lovelock
Lagrangian vanishes, ${\cal L}^N=0$, for $N\geq3$), it is only the GB term that could
have a non-trivial contribution, through a scalar field coupling, on the $4$-dimensional 
dynamics. 

We have demonstrated that a pure scalar-GB theory naturally supports de Sitter
solutions with a natural exit mechanism to a linearly expanding -- Milne -- phase. 
The quadratic coupling function, although a special choice, is in fact the only
one that allows for inflationary, de Sitter-type solutions to emerge \cite{DGK2};
in addition, it shares several features
with the one emerging in the context of the heterotic superstring effective action and
was shown in the past to support similar type of singularity-free cosmological
solutions \cite{KRT}. Our simplifying approach leads to analytical, elegant inflationary
solutions with a bounded potential and free of fine-tunings. 
Important questions that still need to be answered are how, and at which
point in time, the Ricci scalar and the presence of additional matter affect the
dynamics of the model -- for this, a more involved numerical analysis is clearly
necessary. Also, one should investigate whether this type of
inflation gives a perturbation spectrum in accordance to observations.
Note, however, that all the aforementioned questions
will be relevant not only to our model but to any string-inspired or
generalised gravitational theory that contains the Ricci scalar and GB term.
We plan to report on these questions in a future work.

\vskip -0.8cm

\section{Acknowledgments}

This research has been co-financed by the European
Union (European Social Fund - ESF) and Greek national funds through the
Operational Program ``Education and Lifelong Learning'' of the National
Strategic Reference Framework (NSRF) - Research Funding Program: 
``THALIS. Investing in the society of knowledge through the European
Social Fund''. Part of this work was supported by the COST Action MP1210
``The String Theory Universe''.



\end{document}